\documentclass[submission,copyright,creativecommons]{eptcs}
\usepackage{amsmath}
\usepackage{bussproofs}
\usepackage{caption}
\usepackage{pdfsync}
\usepackage{enumerate}
\usepackage[T1]{fontenc}
\usepackage{textcomp}

\usepackage[pdftex]{graphicx} 

\DeclareMathAlphabet{\mathcal}{OMS}{cmsy}{m}{n}

\title{Channels as Objects \\
in Concurrent Object-Oriented Programming}
\author{Joana Campos
\institute{Lasige and Department of Informatics\\
University of Lisbon, Portugal}
\and
Vasco T. Vasconcelos
\institute{Lasige and Department of Informatics\\
University of Lisbon, Portugal}
}
\def\titlerunning{Channels as Objects}
\def\authorrunning{Joana Campos, and Vasco T. Vasconcelos}

\usepackage{listings}
\usepackage{xspace}

\newcommand{\ea}{\emph{et al.}\xspace}

\newcommand*{\set}[1]{\{{#1}\}}

\let\subset\subseteq
\let\phi\varphi

\newcommand{\bnf}{\;::=\;}
\newcommand{\alt}{\;\;|\;\;}
\newcommand\classterm{\mkterm{class}}

\newcommand*{\class}[4]{\mkterm{class}~{#1}~\{#2;#3;#4\}}

\newcommand*{\method}[4]{{#1}~{#2}(#3)~\{#4\}}

\newcommand*{\mkterm}[1]{\mathsf{#1}}

\newcommand\this{\mkterm{this}}

\newcommand\fieldsterm{\mkterm{fields}}

\newcommand*{\objecttype}[2]{{#1}[#2]}

\newcommand*{\methcal}[3]{{#1}.{#2}({#3})}

\newcommand*{\seq}[2]{{#1};{#2}}
\newcommand*{\assign}[2]{{#1}~=~{#2}}
\newcommand*{\new}[1]{\mkterm{new}~{#1}()}
\newcommand{\reduces}{\longrightarrow}

\newcommand*{\subs}[2]{\{^{#1}\!/_{#2}\}}
\newcommand*{\state}[2]{({#1};~{#2})}

\newcommand*{\branch}[3]{\set{#1:#2}_{#3}}

\newcommand{\whileterm}{\mkterm{while}}
\newcommand*{\while}[2]{\whileterm~({#1})~{#2}}

\newcommand{\End}{\mkterm{end}}

\newcommand*{\rectype}[2]{\mu{#1}.{#2}}

\newcommand*{\typedexp}[4]{{#1}\triangleright{#2} : {#3}\triangleleft{#4}}
\def\judgment#1>#2:#3<#4/{\ensuremath{\typedexp{#1}{#2}{#3}{#4}}}
\def\subjudgment#1>#2:#3<#4/{\ensuremath{{#1}\triangleright{#2}\subt{#3}\triangleleft{#4}}}

\newcommand{\typingRuleSkip}{5ex}

\newcommand*\changetype[3]{{#1}\{#2\mapsto #3\}}
\newcommand*\changeval[3]{{#1}\{#2\mapsto #3\}}

\newcommand{\subt}{\mathrel{\mathsf{<:}}}

\newcommand*\rulename[1]{\LeftLabel{\scshape (#1)}}

\newcommand*\axiomname[1]{{\scshape (#1)~}}

\newcommand{\spawnterm}{\mkterm{spawn}}
\newcommand{\spawn}[1]{\spawnterm~#1}

\newcommand{\dom}{\mathit{dom}}

\newcommand*{\field}[2]{{#1}~{#2}}
\newcommand\unit{\mkterm{unit}}
\newcommand{\booleantype}{\mkterm{boolean}}
\newcommand\trueterm{\mkterm{true}}
\newcommand\falseterm{\mkterm{false}}
 
\newcommand{\lin}{\mkterm{lin}} 
\newcommand{\un}{\mkterm{un}} 
\newcommand{\syncterm}{\mkterm{sync}} 
\newcommand \init{\mkterm{init}}
\newcommand*{\initfunction}[1]{\init ({#1})}
\newcommand{\ifterm}{\mkterm{if}} 
\newcommand{\elseterm}{\mkterm{else}} 
\newcommand*{\ifelse}[3]{\ifterm~({#1})~{#2}~\elseterm~{#3}} %

\newcommand*\change[2]{#1\mapsto #2}

\newcommand\insyncterm{\mkterm{insync}}
\newcommand*{\insync}[2]{\insyncterm~{#1}~{#2}}
\newcommand\usageterm{\mkterm{usage}}
\newcommand*\choicem[2]{\langle #1+#2\rangle} 
\newcommand*{\branchm}[3]{\set{#1;#2}_{#3}}

\newcommand*{\myhentry}[3]{({#1},{#2},{#3})}

\newcommand\lockterm{\mkterm{lock}}
\newcommand\methodsterm{\mkterm{methods}}

\newcommand*{\myhadd}[2]{{#1},{#2}}
\newcommand\uninitterm{\mkterm{uninit_{C[u]}}}
\newcommand\mool{\textsc{Mool}}

\newcommand\thenterm{\mkterm{then}}

\def\judgmentu#1>#2#3<#4/{\ensuremath{\typedusage{#1}{#2}{#3}{#4}}}
\newcommand*{\typedusage}[4]{{#1}\triangleright{_#2}~{#3} \triangleleft{#4}}

\begin{document}

\maketitle

\begin{abstract}
  There is often a sort of a protocol associated to each class,
  stating when and how certain methods should be called. Given that
  this protocol is, if at all, described in the documentation
  accompanying the class, current mainstream object-oriented languages 
  cannot provide for the verification of client code adherence  
  against the sought class behaviour.
% However, clients often must know when/how to call methods
% in order for objects to work properly.
%
  We have defined a class-based concurrent object-oriented language
  that formalises such protocols in the form of \emph{usage types}. 
  Usage types are attached to class definitions, allowing
  for the specification of (1)~the available methods, (2)~the tests clients
  must perform on the result of methods, and (3)~the object status ---
  linear or shared --- all of which depend on the object's state. Our
  work extends the recent approach on modular session types by
  eliminating channel operations, and defining the method call as the
  single communication primitive in both sequential and concurrent
  settings. In contrast to previous works, we define a single category
  for objects, instead of distinct categories for linear and for
  shared objects, and let linear objects evolve into shared ones. We
  introduce a standard $\syncterm$ qualifier to prevent thread
  interference in certain operations on shared objects. We formalise
  the language syntax, the operational semantics, and a type system
  that enforces by static typing that methods are called only when
  available, and by a single client if so specified in the usage
  type. We illustrate the language via a complete example.
\end{abstract}

%%% Local Variables: 
%%% mode: latex
%%% TeX-master: "main"
%%% End: 

\section{Motivation}
\label{sec:motivation}

Today's mainstream object-oriented languages statically check code
against object interfaces that describe method signatures only.
% (standard) type and initialisation errors.
There are often semantic restrictions in the implementation of classes
that impose particular sequences of legal method calls and aliasing
restrictions, which clients must observe in order for objects to work
properly. Usually, these usage protocols are only described in
informal documentation, and hence any disregard cannot be statically
detected, revealing itself as a runtime exception (in languages
equipped with built-in support to handle exceptional events) or, worse
still, in memory corruption and unexpected behavior.

In this paper, we present our work on the language $\mool$, a mini
object-oriented language in a Java-like style with support for
concurrency, that allows programmers to specify usage protocols. The
language includes a simple concurrency mechanism for thread
spawning. Class protocols are formalised as \emph{usage types} that
specify (1)~the available methods, (2)~the tests clients must perform
on the values returned by methods, and (3)~the existence of aliasing
restrictions, all of which depend on the object's state.

Aliasing makes it difficult for a client to know if it is complying
with the protocol since multiple references may exist to the same
object, and may alter its state. In our approach, we define a single
category for objects, as opposed to distinct categories for linear and
for unrestricted (or shared) objects. We let the current status of an
object to be governed by its type, allowing linear objects to evolve
into shared ones
(\textit{cf.}~\cite{VasconcelosV:session-types-functional,
  vasconcelos:fundamental-sessions}). The opposite is not possible, as
we do not keep track of the number of references to a given object. We
propose that the programmer explicitly introduces annotations in the
usage descriptor, in order to distinguish between \lstinline|lin| and
\lstinline|un| object states. The \lstinline|lin| qualifier describes the
status of an object that can be referenced in exactly one thread
object. The \lstinline|un| qualifier stands for unrestricted, or
shared, and governs the status of an object that can be referenced in
multiple threads. The type system further maintains that unrestricted objects
do not contain more restrictive linear attributes.

% The modularity on the foregoing
% work~\cite{gay.vasconcelos.etal_modular-session-types} comes from
% partitioning the protocol implementation into several methods, as
% opposed to other approaches in the context of object-oriented
% languages in which protocols are implemented in a single method
% body~\cite{DezaniCiancagliniM:sestoo,DezaniCiancagliniM:disool}.

In $\mool$, we adopt an approach similar to the one of Gay
\ea~\cite{gay.vasconcelos.etal_modular-session-types} defining a
global usage specification of the available methods. We spread the
implementation of a usage type over separate methods and classes,
following a modular programming principle. This means that we also have
to handle \textit{non-uniform objects}, that is, objects that
dynamically change the set of available
methods. Nierstrasz~\cite{nierstrasz:regular-types} was the first to
study the behaviour of non-uniform, or active, objects in concurrent
systems.
We extend Gay's work in two ways. (1) The foregoing work treats
communication channels shared by different threads as objects, by
hiding channel primitive operations in an API from where clients can
call methods. In $\mool$, we eliminate channels and define a
programming language that relies on a simpler communication model --
message passing in the form of method calls, both in the sequential
and in the concurrent setting. (2) The foregoing work deals with
linear types only.  In $\mool$, we deal with linear types as well as
shared ones, treating them in a unified category.

Setting a single category for objects introduces an additional
complication in the control of concurrent access to shared objects.
%  Because we do not keep track of the number of references
% existing to a shared object, we do not allow an object with an
% unrestricted type to go back to being linear. Thus, some additional
% mechanism must be introduced.
Since our main focus are linear objects, we adopt a standard and straightforward
solution similar to the \textit{synchronized} methods used in Java in order 
to enforce serialised access to methods that manipulate shared data, and should 
not be executed by two threads simultaneously.

Summing up the main contributions of our approach regarding session
types for objects: 
\begin{itemize}
\item In contrast to other
  works~\cite{DezaniCiancagliniM:sestoo,DezaniCiancagliniM:disool,gay.vasconcelos.etal_modular-session-types},
  we elect method invocation as the only communication model in both
  concurrent and sequential programming;
\item We annotate classes with a usage descriptor to structure method
  invocation, and we enhance it with \lstinline|lin|/\lstinline|un|
  qualifiers for aliasing control, thus defining a single category for
  objects that may evolve from a linear status into an unrestricted
  one;
\item In contrast to other works on session
  types~\cite{gay.vasconcelos.etal_modular-session-types,HondaK:lanptd},
  we replace the shared channel primitive (on which sessions are
  started) by a conventional \lstinline|sync| method modifier, enabling
  multiple threads to independently work on shared operations without
  interfering with each other.
\end{itemize}

More examples, detailed explanations of the core language and of its
implementation in a prototype compiler, as well as a more complete
overview of the literature, are given in the MSc thesis of the first
author~\cite{campos:linear-shared-objects-concurrent-programming}.
The remaining sections are organised as follows:
Section~\ref{sec:example} introduces the language via an example;
Section~\ref{sec:language} gives an overview of the language
technicalities; and Section~\ref{sec:conclusion} contains additional
discussion on related work, as well as an outline of future work.

% This is a short version of the thesis "Linear and Shared Objects in
% Concurrent Programming" submitted in partial fulfilment of the
% requirements of the Faculty of Sciences of the University of Lisbon
% for the MSc Degree in Informatics Engineering.  

%%% Local Variables: 
%%% mode: latex
%%% TeX-master: "main"
%%% End: 

\section{The Auction System}
\label{sec:example}

Our auction system, adapted from 
\cite{VallecilloVasconcelosEtal:typing-behavior-objects,
VasconcelosV:session-types-functional}, features three kinds of participants: 
the auctioneer, the sellers and the bidders. Sellers sell items for a minimum 
price. Bidders place bids in order to buy some item for the best possible 
price. The auctioneer controls these interactions.

\begin{figure}[ht]	
	\begin{minipage}{0.48\linewidth}
		\includegraphics[width=\textwidth]{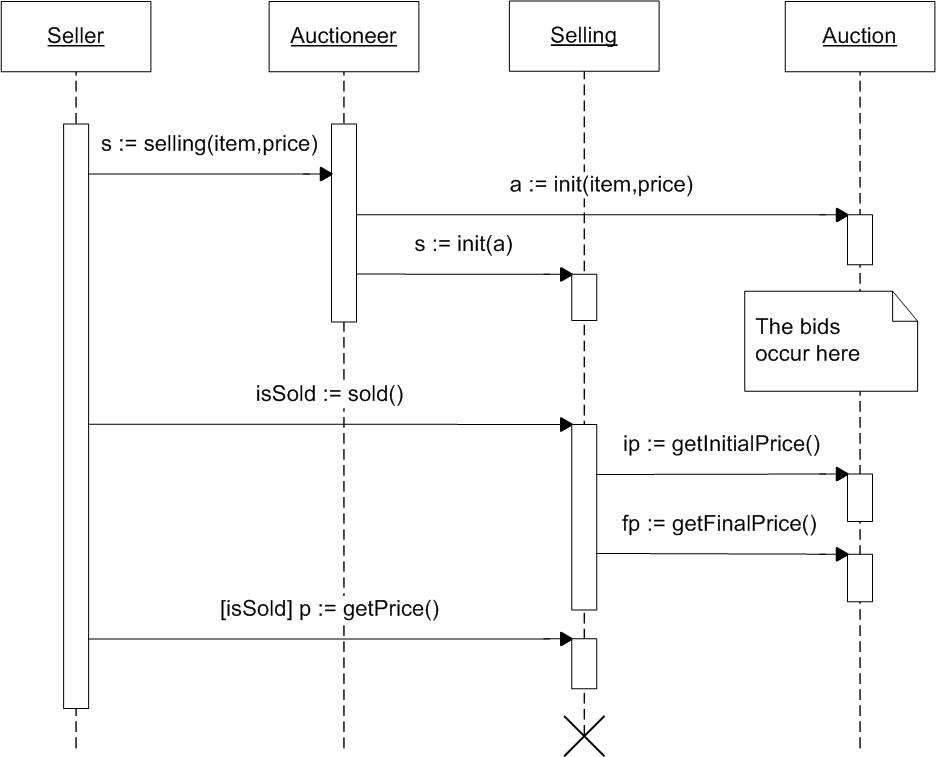}		
		\caption{The scenario for a seller}
		\label{fig:seller-diagram}
	\end{minipage}
	\hspace{0.25cm} 
	\begin{minipage}{0.48\linewidth}		
		\includegraphics[width=\textwidth]{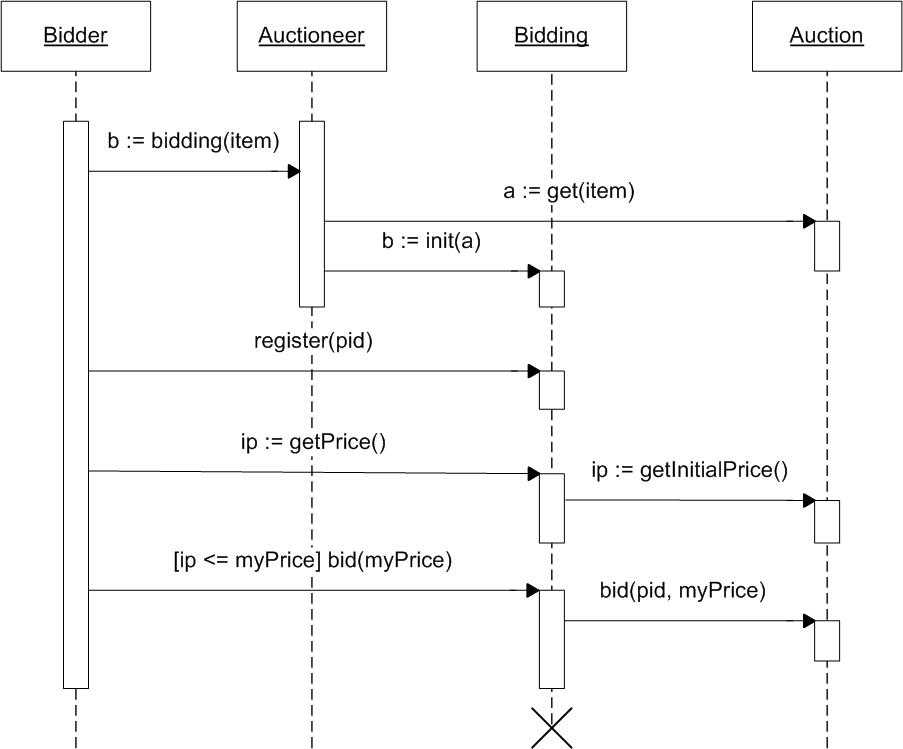}		
		\caption{The scenario for a bidder}
		\label{fig:bidder-diagram}
	\end{minipage}
\end{figure}

The system is best described by the UML sequence diagrams in 
Figures~\ref{fig:seller-diagram} and~\ref{fig:bidder-diagram}, modelling the two main 
scenarios through a sequence of messages exchanged between objects. The first diagram 
describes the scenario for a seller, while the second one describes the scenario for 
a bidder. We are mainly interested in visualising the interaction between the 
different players; we do not represent concurrent communication, even though it 
should not be hard to imagine several seller and bidder threads concurrently making 
requests to the same auctioneer object. To lighten the representation, we omit the 
usual dashed open-arrowed line at the end of an activation box indicating the return 
from a message and its result. Instead, we simply annotate sent messages with the 
returned value, if any. A conditional message is shown by preceding the message by a 
conditional clause in square brackets. The message is only sent if the clause 
evaluates to $\trueterm$.

The first diagram (Figure~\ref{fig:seller-diagram}) depicts how a \lstinline|Seller| 
initiates the interaction with the \lstinline|Auctioneer| indicating the item to 
auction and its price. The \lstinline|Auctioneer|, after creating an 
\lstinline|Auction| object, where the bids for the item being sold are going to be 
placed, delegates the service to a new \lstinline|Selling| object. Once the auction 
is set, the auctioneer can start receiving bids for that item. We signal with a note 
the moment when the two scenarios interweave. In the interaction initiated by a 
\lstinline|Bidder| (Figure~\ref{fig:bidder-diagram}), the  \lstinline|Auctioneer| 
receives a request for the item searched and, if the item is being auctioned, it 
delegates the service to the \lstinline|Bidding| object. We can see that a 
\lstinline|Bidder| holds its own \lstinline|Bidding| object from which it can obtain 
the item initial price as defined by the \lstinline|Seller|. Based on the returned 
value, the \lstinline|Bidder| decides whether to make a bid by calling the 
appropriate method on the \lstinline|Bidding| object. Back to 
the seller scenario, if the sale is successful, the \lstinline|Selling| object 
allows a \lstinline|Seller| to obtain the sale final price by invoking method 
\lstinline|getFinalPrice|.

Both scenarios depict a similar pattern, and it is not difficult to
conclude from the diagrams that the usage protocols specified in the
\lstinline|Selling| and \lstinline|Bidding| classes should be
described by linear types. That is the only way we can guarantee that
clients (sellers and bidders) fully obey the specification, by
enforcing that their types are consumed to the end. A fresh
\lstinline|Selling| (and \lstinline|Bidding|) object is created at the
beginning of each selling (and bidding) interaction, and is implicitly
destroyed at the end. We signal object destruction in the diagram to
increase the expressiveness of the representation; the type system
provides crucial information to object deallocation.

%%% Local Variables: 
%%% mode: latex
%%% TeX-master: "main"
%%% End: 

\paragraph{Usage syntactic details} Apart from the usage specification at the 
beginning of each class, our language presents a typical Java-like syntax. Before 
explaining the implementation in detail, we introduce some less obvious syntactic 
details. The usage formalises how clients should use an object of a given class. All 
methods that are not referred in the usage specification are not visible to clients. 
For the same reason, class fields are also private, and cannot be used by other classes.

If a program defines a conventional class 
named \lstinline|C| with methods \lstinline|m1|, \lstinline|m2| and 
\lstinline|m3|, and no usage declaration, our compiler will insert 
\lstinline|usage *{m1 + m2 + m3}| as the class default 
usage type, where each method (\lstinline{m1}, \lstinline{m2} and 
\lstinline{m3}) is always available. Formally, this usage defines 
a recursive branch type of the form \lstinline|mu X.un{m1.X + m2.X + m3.X}|. The 
\lstinline|un| qualifiers are omitted in the examples. A choice between calling one 
of the three available methods is indicated by (+). Because of the particular form of 
the recursive type, calling any of these methods on an instance of  
\lstinline|C| will not change the object state nor the set of available methods.

A typical usage declaration for a linear object is a sequential
composition of available methods. If \lstinline|C| is linear,
\lstinline|usage lin m1; lin m2; lin m3; end;| is a possible usage
declaration. Calling methods in the prescribed order on an instance of
this class changes the object state and the set of available
methods. State \lstinline|end| is an abbreviation for
\lstinline|un {}|. When an object is in this state, it means that the set of
available methods is empty (the usage protocol is finished).

A variant type, denoted by $\langle \ldots + \ldots\rangle$, is indexed by the 
$\trueterm$ and $\falseterm$ values returned by the method to which the variant is 
bound. A client should test the result of the call: if \lstinline|true| is 
returned, the new object state, and the available methods, are to be found on the 
left-hand side of the variant; if \lstinline|false| is returned, it is the right-hand 
side to dictate the object state and available methods. 

\paragraph{Programming with usage types} Consider now the usage
specification in Figure~\ref{fig:auctioneer}. When an object of the
\lstinline|Auctioneer| class is created through the explicit constructor
\lstinline|init|, only one reference exists to it, but then the object
evolves into an unrestricted type, allowing several sellers and
bidders to hold references to the same instance. Notice that we have
defined a recursive (shared) type. Each client object can do one of
two things: (1) it can call method \lstinline|selling| to obtain an
object that provides an implementation of the selling activity on the
auctioneer; or (2) it can call method \lstinline|bidding| and obtain
an object that implements the buying activity on the auctioneer. Then,
it can repeat the interaction all over again: a seller can lower the
price of an item with no bids, and start a new sale; a bidder can bid
a higher price. The type never ends, and this illustrates why, in any
program, we cannot keep track of the number of references to a shared
type.  The return types of these two methods are
\lstinline|Selling[Sold]| and \lstinline|Bidding[Register]|
respectively, because the types of the returned objects have advanced
during the execution of each method body. In the example, the state
\lstinline|Sold| is short for
\lstinline|lin sold; «getPrice; end +  end»;|
(lines~3~and~4 in Figure~\ref{fig:selling}), where~(;) binds 
stronger than (+), and is used for convenience to replace 
the full type above. In declarations, variable types are shortened 
(for space reasons), but their full types are implied.

\begin{figure}[htp]	
	\begin{minipage}{0.45\linewidth}		
		\input{fig-auctioneer}		
		\hspace{4cm}
		\input{fig-seller}
	\end{minipage}
	\hspace{0.6cm}
	\begin{minipage}{0.5\linewidth}
		\input{fig-auction}
		\input{fig-selling}
		\input{fig-main}
	\end{minipage}		 	
\end{figure}

Notice, still in Figure~\ref{fig:auctioneer}, that when a new selling request 
is made, a new \lstinline|Auction| object is created (line~10). This instance is 
then added to the \lstinline|AuctionMap| object (whose class we omit), where the 
\lstinline|Auctioneer| keeps all the auctions (line~12), and is passed 
to the constructors of both the \lstinline|Selling| and \lstinline|Bidding| 
objects. In the \lstinline|selling| method, the reference to the \lstinline|Auction| is 
created locally within the method body and assigned to the variable named \lstinline|a| 
(line~10), while in the \lstinline|bidding| method, it must be fetched from the 
\lstinline|AuctionMap| object (line~20). It is through reading and writing to this 
shared \lstinline|Auction| object that the protocol takes place. 

The usage declaration in the \lstinline|Auction| class (Figure~\ref{fig:auction}) 
shows another example of a recursive type in a shared object. Notice also the 
\lstinline|sync| method qualifier (line~10) that is used to control concurrent bids 
made by separate \lstinline|Bidder| threads. This annotation also 
qualifies the \lstinline|put| and \lstinline|get| operations in the 
\lstinline|AuctionMap| class (omitted).

Figures~\ref{fig:seller} and~\ref{fig:selling} implement two linear
types. The usage declaration of the \lstinline|Seller| class 
(Figure~\ref{fig:seller}) is an abbreviation for the nested composition of 
branch types \lstinline|lin{init; lin{run; un{}}}|. An example
of a variant type is provided in the \lstinline|Selling| class (line~4 in 
Figure~\ref{fig:selling}). A variant type
is always linear, so the redundant \lstinline|lin| qualifier can be omitted. 
This type requires that the client evaluates the returned boolean 
value of the \lstinline|sold| method in order to determine the next available 
method: if the value evaluates to \lstinline|true|, then the caller can 
obtain the price via \lstinline{getPrice} (because the item was sold), 
otherwise the interaction ends.

The \lstinline|lowerPrice| method in the \lstinline|Seller| class
(Figure~\ref{fig:seller}) is not referred in the usage specification (and 
is thus omitted from the object type). Our type system 
ensures that only methods in the usage type are visible to 
clients~(\textit{cf.}~\cite{gay.vasconcelos.etal_modular-session-types}). 
All methods not specified in the usage type can be accessed from their own classes, 
but never alter the type (otherwise client views of the object could become 
inconsistent).

Finally, the \lstinline|Main| class in Figure~\ref{fig:main} creates an 
\lstinline|Auctioneer| object that controls the auction, and
spawns a separate thread for a \lstinline|Seller| and two
\lstinline|Bidder| objects, using a Java-like technique for thread
creation.

%%% Local Variables: 
%%% mode: latex
%%% TeX-master: "main"
%%% End: 

\section{The Core Language}
\label{sec:language}

Most of the technicalities that we now present are based on the core
language from modular session types by 
Gay~$\ea$~\cite{gay.vasconcelos.etal_modular-session-types}, and on the
linear type system presented by
Vasconcelos~\cite{vasconcelos:fundamental-sessions}.  Inspired by
these approaches to sessions types, our main challenge was the attempt
to formalise the convergence of channels and objects in a simple
concurrent object-oriented language.

\paragraph{Syntax}

In $\mool$, we distinguish between the user syntax and the runtime syntax.
The former is the programmer's language, and is defined in 
Figure~\ref{fig:syntax}; the latter is only required by the 
operational semantics, and appears in Figure~\ref{fig:syntax-ext}.

\begin{figure}[t]
  \begin{align*}
    % CLASS DECLARATIONS
    \textrm{(class declarations)}&& D & \bnf\  \class{C}{u}{\vec F}{\vec M}     
    \\
    % FIELD DECLARATIONS
    \textrm{(field declarations)}&& F & \bnf\ 
    \field{t}{f}
    \\      
    % METHOD DECLARATIONS
    \textrm{(method declarations)}&& M & \bnf\ 
    s~\method{t}{m}{t~x}{e}
    \\    
    % METHOD QUALIFIERS
    \textrm{(method qualifiers)}&& s & \bnf\ 
    \varepsilon \alt \syncterm   
    \\       
    % TYPES
    \textrm{(types)}&& t & \bnf\ \unit \alt \booleantype \alt 
    \objecttype C u 
    \\ 
    % VALUES
    \textrm{(values)}&& v & \bnf\ \unit \alt \trueterm \alt \falseterm \alt o       
    \\                       
    % EXPRESSIONS
    \textrm{(expressions)}&& e & \bnf\ v \alt x \alt o.f \\
    &&& \alt \seq e e \alt \assign{o.f}{e} \\
    &&& \alt \new C \alt \methcal{o}{m}{e} \alt \methcal{o.f}{m}{e}\\     
    &&& \alt \ifelse{e}{e}{e} \alt \while{e}{e} \\
    &&& \alt \spawn{e} 
    \\       
    % USAGE TYPES
    \textrm{(class usage types)}&& u & \bnf\  
    q~\branchm{m_i}{u_i}{i\in I} \alt \choicem{u}{u} \alt 
    \\
    &&& \alt X \alt \mu X.u
    \\  
    % USAGE QUALIFIERS
    \textrm{(type qualifiers)}&& q & \bnf\ 
    \lin \alt \un     
  \end{align*} 
  \caption{User syntax}
  \label{fig:syntax}
\end{figure}

%%% $Id: fig-syntax.tex,v 1.11 2006/12/13 17:04:24 vv Exp $
%%% Local Variables: 
%%% mode: latex
%%% TeX-master: "main"
%%% End: 

\begin{figure}[t]
  \begin{align*}        
    % VALUES
    \textrm{(values)}&& v & \bnf\ 
    \ldots \alt \uninitterm
    \\
    % EXPRESSIONS
    \textrm{(expressions)}&& e & \bnf\ \ldots \alt \insync{o}{e}  
    \\  
    % CONTEXTS
    \textrm{(contexts)}&& \mathcal{E} & \bnf\ [\_] \alt
    \seq{\mathcal E}{e} \alt \assign{o.f}{\mathcal E} \\
    &&& \alt \methcal{o}{m}{\mathcal{E}} \alt
			\methcal{o.f}{m}{\mathcal{E}}  \\ 
    &&& \alt \ifelse{\mathcal E}{e}{e} \alt \while{\mathcal E}{e}  
    \\   
    % LOCKS
    \textrm{(lock flags)}&& l & \bnf\ 0 \alt \ 1
    \\          
    % OBJECT RECORDS
    \textrm{(object records)}&& R & \bnf\ 
    \myhentry{t}{l}{\vec f=\vec v}
    \\        
    % HEAPS
    \textrm{(heaps)}&& h & \bnf\ \emptyset \alt \myhadd{h}{o=R}
    \\
    % STATES     
    \textrm{(states)}&& S & \bnf\ \state{h}{e_1|\dots|e_n}    
    %\\
    % ENVIRONMENTS
    %\textrm{(environments)}&& \Gamma & \bnf\ \emptyset \alt \Gamma,r:t     
    %\\
    % VISITED USAGE TYPES
    %\textrm{(visited usage types)}&& \Theta & \bnf \emptyset \alt \Theta   
  \end{align*}
  \caption{Runtime syntax}
  \label{fig:syntax-ext}  
\end{figure}

%%% $Id: fig-syntax.tex,v 1.11 2006/12/13 17:04:24 vv Exp $
%%% Local Variables: 
%%% mode: latex
%%% TeX-master: "main"
%%% End: 

In order to simplify both the operational semantics and the
type system, we introduce some restrictions on the syntax of specific
constructs: (1) an assignment expression is only defined on fields (not on
parameters); (2) a method call is only defined on a field of an object
reference (not on an arbitrary expression); (3) it follows from the
above that calling a method on a parameter requires assigning it first
to a field; and (4) methods accept exactly one parameter.

We assume that class identifiers in a sequence of declarations $\vec D$ are all 
distinct, and that the set of method and field names declared in each class contains 
only distinct names as well. Object references $o$ include the keyword $\this$, which 
references the current instance. In the examples, field accesses omit the prefix 
$\this$; the compiler can insert it when needed. Method modifiers, which include the  
$\syncterm$ keyword, are optional, which we indicate in the syntax by the empty string~$\varepsilon$. 

The $\mool$ syntax defines non-object and object types. Non-object
types include the primitive $\unit$ and $\booleantype$ types. The
$\unit$ type has the single value $\unit$, while the $\booleantype$
type comprises two values: $\trueterm$ and $\falseterm$. The type of
an object is $\objecttype C u$ in the style of modular session
types~\cite{gay.vasconcelos.etal_modular-session-types}, describing an
object of a class named~$C$ at usage type~$u$. From a client class
perspective, it prescribes the structure of method invocation: in
which order/how methods should be called.

The $\mool$ expressions are standard in object-oriented languages. We have defined 
(in order of appearance) values, parameters, fields, the sequential expression 
composition, assignment to fields, object creation, the self-call, the method call 
on fields, control flow expressions, and thread creation using the $\spawnterm$ 
primitive. The form of the $\spawnterm$ expression means that the inner expression 
$e$ is evaluated in a newly created thread.

In the core language, only the abstract syntax of usage types is
defined, omitting the \lstinline|usage| and \lstinline|where| clauses,
and the branch choice (+) and recursive (*) operators that have been
introduced in the examples.  However, it is not difficult to translate
from the syntax of the examples into the core language syntax. For
example, the recursive type 
\lstinline|usage lin init; *{selling + bidding};| in the 
\lstinline|Auctioneer| class takes the following configuration 
in the core language:
$\lin\branchm{\textsf{\small init}}{\mu X. \un{\{\textsf{\small
      selling}.X, \textsf{\small bidding}.X\}}}{}$.

Branch types are denoted by $\branchm{m_i}{u_i}{i\in I}$, defining
available methods $m_i$ and their continuations $u_i$. Variant types
are denoted by $\choicem{u}{u}$, and are indexed by the two
$\booleantype$ values, $\trueterm$ first. In order to resolve the type, the caller must
perform a test on the result of the preceding method. For simplicity,
we use binary-only variants as in
typestates~\cite{StromYemini:typestates}, but more generous variants
using enumerations can be found in the
literature~\cite{gay.vasconcelos.etal_modular-session-types}. Recursive
types are indicated in the syntax by $\mu X.u$, and need to be
\emph{contractive}, which means that no subexpression of the form
$\rectype{X_1}{\cdots\rectype{X_n}{X_1}}$ is
allowed~\cite{gay.vasconcelos.etal_modular-session-types}. These types
are treated in an equi-recursive discipline, which means that we
regard $\mu X.u$ and its unrolling $u\subs{\rectype{X}{u}}{X}$ as
equal types.

A usage branch type is annotated with a qualifier $q$ for aliasing control. Recall 
that we define a single category for objects, as opposed to distinct categories for 
linear and for unrestricted objects, allowing linear objects to evolve into a status 
where they can be shared by multiple clients, but not the opposite  
(\emph{cf.}~\cite{VasconcelosV:session-types-functional, 
vasconcelos:fundamental-sessions}). 

The type system, that we introduce later, imposes some
further restrictions on usage types, namely that: (1) the initial
class usage type must be a branch, because a variant type is bound to
the result of a method call; (2) in a variant type configuration
$\choicem{u}{u}$, each $u$ corresponds to a $\lin$-qualified variant
(and that is why the qualifier is always omitted); (3) an object
final status is always unrestricted, whether because it evolves into
an $\End$ state (with an empty set of alternatives), or because, being
recursive, it never ends, as illustrated in the examples; 
and (4) unrestricted objects may not contain linear attributes.

The runtime syntax in Figure~\ref{fig:syntax-ext} introduces additional elements 
required by the operational semantics, but not available in the 
user syntax. Values include $\uninitterm$, which is used by our compiler to initialise 
fields of type $C[u]$ when an object is created. Expressions are extended with the 
$\insyncterm$ expression in the style of Flanagan and Abadi~\cite{flanaganA99}, 
denoting that the subexpression $e$ is currently being evaluated while the lock is 
held on object $o$. 

Evaluation contexts are denoted by $\mathcal{E}$, and are expressions
with one hole.  Contexts specify the order in which expressions are
evaluated, defining where reduction rules can be applied. A heap is
viewed as a map (a partial function of finite domain) from object
identifiers~$o$ into records $R$. The operation $\myhadd{h}{o=R}$ adds
an entry to the heap $h$, if $o$ does not yet exist, and is considered
to be associative and commutative, which means that a heap is an
unordered set of bindings
\cite{gay.vasconcelos.etal_modular-session-types}. Object records are
instances of usage-typed classes and are represented by the triple
$\myhentry{t}{l}{\vec f=\vec v}$, where $t$ is the object current
type, $l$ represents the object lock flag, and $\change{\vec f}{\vec
  v}$ is a mapping from field identifiers to their values. The
\emph{lock} can be either in a locked or unlocked state, denoted by
the 1 and 0 flag, respectively. The record $\myhentry{\_}{1}{\_}$
means that some thread currently holds the lock associated with the
current instance.  Initially, an object is created with the flag set
to 0. States consist of two components: a heap and a parallel
composition of expressions sharing the heap.

%%% Local Variables: 
%%% mode: latex
%%% TeX-master: "main"
%%% End: 

\paragraph{Operational Semantics} 

Figure~\ref{fig:reduction-states} defines the 
reduction rules for states, taking the form of a parallel composition of expressions: 
\begin{center}
$\state{h}{e_1|\dots|e_n}  \reduces \state{h'}{e_1'|\dots|e_m'} $
\end{center}

\begin{figure}[t]
\centering\setlength\lineskip{\typingRuleSkip}
\AxiomC{$\state{h}{e_1|\dots|e|\dots|e_n} \reduces\state{h'}{e_1|\dots|e'|\dots|e_n}$}
\rulename{R-Context}
\UnaryInfC{$\state{h}{e_1|\dots|\mathcal{E}[e]|\dots|e_n}
\reduces\state{h'}{e_1|\dots|\mathcal{E}[e']|\dots|e_n}$}
\hfil\\
\DisplayProof
\axiomname{R-Spawn}
$\state{h}{e_1|\dots|\mathcal{E}[\spawn{e}]|\dots|e_n}
\reduces\state{h}{e_1|\dots|\mathcal{E}[\unit]|e|\dots|e_n}$
%\hfil\\
%\hfil\\
\caption{Reduction rules for states}
\label{fig:reduction-states}
\end{figure}
%%% $Id$

%%% Local Variables: 
%%% mode: latex
%%% TeX-master: "main"
%%% End: 

\textsc{R-Context} is standard, defining which expression can be evaluated next in 
the program execution. It is the rule that invokes all the other reduction rules for 
expressions (see Figure~\ref{fig:reduction-exps}). \textsc{R-Spawn} detaches 
the expression to a new thread running in parallel. After $\spawnterm$, the original 
thread proceeds to evaluate its next instruction. The value of the $\spawnterm$ 
expression is $\unit$, as the result of evaluating $e$ is not transferred back to the 
original thread.

The rules for expressions in Figure~\ref{fig:reduction-exps} take a similar form, 
with the state containing only one expression:
\begin{center}
$\state{h}{e} \reduces \state{h'}{e'}$
\end{center}

The rules use in their definitions the predicates defined in 
Figure~\ref{fig:predicates}. Predicate $q(v)$ determines the type status 
(linear or unrestricted) given a value $v$. If $v$ is a value of a primitive type, 
then its status is always unrestricted. If $v$ is an object, then its status 
depends on its current type, which can be fetch from the usage component of the 
record associated with~$o$~in~heap~$h$. A variant type is always linear, and the 
status of a branch type is defined by the programmer. Predicate $q(t)$ follows a 
similar pattern, with the difference that it takes a type as its argument. Finally, 
function $\initfunction{t}$ takes a type as its argument and returns the 
type default value.

\begin{figure}[t]
\centering\setlength\lineskip{\typingRuleSkip}
\begin{tabular}{p{8cm}}
  {\bf{Predicates $\lin(v)$ and $\un(v)$ with $q \bnf\ \lin \alt \un$}} 
  \begin{itemize}
    \item if $v$ = $\unit, \trueterm, \falseterm$ then $\un(v)$
    \item if $v$ = $o$ and $h(o).\usageterm$ = $\choicem{\_}{\_}$ then $\lin(v)$
    \item if $v$ = $o$ and $h(o).\usageterm$ =  $q~\lbrace \dots \rbrace$ then $q(v)$
  \end{itemize} 
  {\bf{Predicates $\lin(t)$ and $\un(t)$ with $q \bnf\ \lin \alt \un$}} 
  \begin{itemize}
     \item if $t$ = $\unit, \booleantype$ then $\un(t)$
     \item if $t$ = $C[\choicem{\_}{\_}]$ then $\lin(t)$
     \item if $t$ = $C[q~\lbrace \dots \rbrace]$ then $q(t)$
  \end{itemize} 
  {\bf{Function $\init(t)$}}
  \begin{itemize}
     \item $\initfunction{\booleantype} = \falseterm$
     \item $\initfunction{\unit} = \unit$
     \item $\initfunction{\objecttype{C}{u}} = \uninitterm$
  \end{itemize}
\end{tabular}
\caption{Auxiliary functions for values and types}
\label{fig:predicates}
\end{figure}

\textsc{R-LinField} and  \textsc{R-UnField} extract the value of the field from an 
object in the heap and determine whether the value belongs to an unrestricted or a 
linear type. After access, the value of an unrestricted field is $v$ as in Java, but 
the value of a linear field must be $\unit$ to ensure that only one reference to an 
object with a linear type exists at any given moment. 

\begin{figure}[t]
\centering\setlength\lineskip{\typingRuleSkip}
\AxiomC{$h(o).f = v$}
\AxiomC{$\lin(v)$}
\rulename{R-LinField}
\BinaryInfC{$\state{h}{o.f}\reduces\state{\changeval{h}{o.f}{\unit}}{v}$}
\DisplayProof
\hfil
\AxiomC{$h(o).f = v$}
\AxiomC{$\un(v)$}
\rulename{R-UnField}
\BinaryInfC{$\state{h}{o.f}\reduces\state{h}{v}$}
\DisplayProof
\hfil
\hfil\\
\axiomname{R-Seq}
$\state{h}{\seq ve} \reduces \state{h}{e}$
\hfil
\axiomname{R-Assign}
$\state{h}{\assign{o.f}{v}}\reduces\state{\changeval{h}{o.f}{v}}{\unit}$
\hfil
\AxiomC{$o$ fresh}
\AxiomC{$C.\fieldsterm = \vec t~\vec f$}
\rulename{R-New}
\BinaryInfC{$\state{h}{\new C}\reduces
\state{\myhadd{h}{o=\myhentry{C}{C.\usageterm}{0}{\vec f = \initfunction{\vec t}}}}{o}$}
\DisplayProof
\hfil
\AxiomC{$\method{\_}{m}{\_\: x}{e}\in (h(o).\classterm).\methodsterm$}
\rulename{R-SelfCall}
\UnaryInfC{$\state{h}{\methcal{o}{m}{v}} \reduces
\state{h}{e\subs{o}{\this}\subs{v}{x}}$}
\DisplayProof
\AxiomC{$m \in h(o).f.\usageterm$}
\AxiomC{$\method{\_}{m}{\_\: x}{e}\in (h(o).f.\classterm).\methodsterm$}
\rulename{R-Call}
\BinaryInfC{$\state{h}{\methcal{o.f}{m}{v}} \reduces
\state{h}{e\subs{o.f}{\this}\subs{v}{x}}$}
\DisplayProof
\AxiomC{$h(o).f.\lockterm = 0$}
\AxiomC{$m \in h(o).f.\usageterm$}
\noLine
\BinaryInfC{$\syncterm~\method{\_}{m}{\_\: x}{e}\in 
   (h(o).f.\classterm).\methodsterm$}
\rulename{R-SCall}
\UnaryInfC{$\state{h}{\methcal{o.f}{m}{v}} \reduces \hfill \state{\changeval{h}{(o).f.\lockterm}{1}}{\insync{o}{e\subs{o.f}{\this}
\subs{v}{x}}}$}
\DisplayProof
\hfill
\axiomname{R-InSync}
$\state{h}{\insync{o}{v}}
\reduces \state{\changeval{h}{(o).\lockterm}{0}}{v}$
\hfill\\
\hfil\\
\hfil
\axiomname{R-IfTrue}
$\state{h}{\ifelse{\trueterm}{e'}{e''}} 
\reduces \state{h}{e'}$
\hfil
\axiomname{R-IfFalse}
$\state{h}{\ifelse{\falseterm}{e'}{e''}} 
\reduces \state{h}{e''}$
\hfil\\
\hfil\\
\axiomname{R-While}
$\state{h}{\while{e}{e'}} \reduces \state{h}{\ifelse{e}{\{e';\while{e}{e'}\}}{\unit}}$
\hfil
\caption{Reduction rules for expressions}
\label{fig:reduction-exps}
\end{figure}
%%% $Id$

%%% Local Variables: 
%%% mode: latex
%%% TeX-master: "main"
%%% End: 

\textsc{R-Seq} reduces the result to the second part of the sequence of expressions, 
discarding the first part only after it has become a value. \textsc{R-Assign} updates 
the value of the field with value $v$. The value of the entire expression needs to be 
$\unit$ (as opposed to $v$), and this is again a linearity constraint. \textsc{R-New} 
generates a fresh object identifier and obtains the field names and types from the 
set of the class fields. Notice that a new object record, indexed by the object 
identifier $o$, is added to the heap, that the object starts in an unlocked 
state~$0$, and that function $\initfunction{t}$ takes a type as its argument and returns 
its default value.

\textsc{R-SelfCall} is for method calls directly on an object reference, which 
typically result from calls on $\this$. These calls do not depend on the usage 
type, unlike rule \textsc{R-Call}. In both rules, if the hypothesis holds, then the 
method body is prepared by replacing occurrences of $\this$ by the receiver heap 
address (so that occurrences of $\this$ within the method body refer to the receiver 
instance) and the actual parameter by the formal one. The resulting expression is the 
method body with the above substitutions that can be further reduced using the 
appropriate rules.

\textsc{R-SCall} is for synchronized calls. The operational semantics proposed by 
Flanagan and Abadi~\cite{flanaganA99} inspired this and \textsc{R-InSync}. The 
new element here is the lock acquired on the receiver object, which  is 
released in \textsc{R-InSync}, by updating the lock component with the flag 0. 
\textsc{R-IfTrue}, \textsc{R-IfFalse} and \textsc{R-While} are standard.

%%% Local Variables: 
%%% mode: latex
%%% TeX-master: "main"
%%% End: 

\newpage
\paragraph{Subtyping} 

The subtyping definition for the $\mool$
language is similar to the one described for the language
in~\cite{gay.vasconcelos.etal_modular-session-types}. The subtype
usage relation $\subt$, defined by co-induction, includes the 
following rules:

\begin{itemize}
\item If $q\branch{m_i}{u_i}{i\in I} \subt u'$
  then $u' = q\branch{m_j}{u'_j}{j\in J}$ with
  $J\subset I$ and $\forall j\in J, u_j\subt u'_j$
\item If $u \subt \choicem{u'}{u''}$ then $u' \subt u$ or $u'' \subt u$ 
\end{itemize}

The branch type is the source of subtyping by allowing that the possible usage 
choices can safely be used in the context where the superusage is expected. 
The proof that the usage relation is a pre-order (i.e. a relation that is reflexive and 
transitive) is not provided here, but can be adapted from~\cite{GaySJ:substp}.

The subtyping usage rule relation is extended to the types of our language as the 
smallest reflexive relation which includes the following: 
$C[u] \subt C[u']$ if $u \subt u'$. This means that implicitly the subtyping relation 
for the two other types in our language is defined as 
$\booleantype \subt \booleantype$ and $\unit \subt \unit$.

%%% Local Variables: 
%%% mode: latex
%%% TeX-master: "main"
%%% End: 

\paragraph{Type System}

The inference rules that define the type system use two typing
environments, $\Theta$~and~$\Gamma$. We consider typing environments
as maps (or partial functions of finite domain) similarly to
heaps. The typing environment~$\Theta$ is a map from usage types $u$
to typing environments~$\Gamma$, and records the field typing
environment associated with usage type $u$. It is used to keep track
of visited usage types, thus preventing cycles in the presence of
recursive types.

The typing assumptions for the environment $\Gamma$ have the form:
\begin{center}
$\Gamma \bnf\ \Sigma \alt \choicem{\Gamma}{\Gamma}$
\end{center}
where $\Sigma$ is a map from fields $o.f$ and parameters $x$ to types $t$. 
The environment $\choicem{\Gamma}{\Gamma}$ represents a pair of maps: the map on the 
left is used for the $\trueterm$ variant, while the map on the right is for the 
$\falseterm$ one.

Below are the typing judgements where these environments are used. The typing 
judgement for usage types is presented on the left, and the one on the right is 
used for expressions:
\begin{center}
$\judgmentu \Theta;\Gamma > C u < \Gamma'/ \qquad \qquad$
$\judgment \Gamma > e : t < \Gamma'/$
\end{center}

The typing judgement for usage types reads: ``A usage type $u$ is valid for a class named $C$ 
starting from field and parameter types in $\Gamma$, the initial environment, and 
ending with field and parameter types in $\Gamma'$, the final environment''. The 
initial and final environments may be different, because object field types 
in~$\Gamma$ may have changed after each method described in $u$ has been 
type-checked.

The typing judgement for expressions follows a similar
pattern. $\Gamma$ is the initial typing environment before
type-checking the expression $e$, and~$\Gamma'$ is the final one, after
associating $e$ with the type $t$. Again, the initial and final
environments may be different, because of side-effects the expression
being checked may cause on the environment $\Gamma$, turning it into
the environment $\Gamma'$. In particular, identifiers may become
unavailable in the case of references to linear objects, and object
types may change as a result of the evaluation of some language
constructs, namely method calls and control flow expressions. Using a
similar notation to the one used to update the heap, if $\Gamma(o.f)$
is defined, i.e., if $o.f \in \dom(\Gamma)$, then
$\changetype{\Gamma}{o.f}{t}$ denotes the environment obtained by
updating to $t$ the type of $o.f$ .

The typing rules defined for $\mool$ have a clear algorithmic configuration, and 
type-checking is modular and performed following a top-down strategy: a program 
checking is conducted by checking each class separately, which in turn conducts the 
checking of each method within the class \emph{in the order in which it appears in 
the specified usage type}. In what follows, rules are presented in the order in which 
they are evaluated by the type-checker (and in the order in which they should be 
read).

\begin{figure}[t]
\centering\setlength\lineskip{\typingRuleSkip}
\AxiomC{$\vdash D_1 \quad \dots \quad \vdash D_n$}
\rulename{T-Program}
\UnaryInfC{$\vdash D_1 \dots D_n$}
\DisplayProof
\hfil
\AxiomC{$\judgmentu \emptyset; \this.\vec f : \vec t > C u < \Sigma / 
   \qquad \un(\Sigma)$}
\rulename{T-Class}
\UnaryInfC{$\vdash\class{C}{u}{\vec t~\vec f}{\_}$}
\DisplayProof
\hfil
\caption{Typing rules for programs}
\label{fig:typing-programs}
\end{figure}

%%% Local Variables: 
%%% mode: latex
%%% TeX-master: "main"
%%% End: 

Figure~\ref{fig:typing-programs} defines rules for typing
programs. When checking that a program is well-formed,
\textsc{T-Program} simply checks that each class defined in it is
well-formed. \textsc{T-Class} constructs the field arguments in the
initial typing environment. Notice that when starting to type-check a
class, the environment $\Theta$ is empty as no usage types have yet
been visited.  \textsc{T-Class} is responsible for calling the rules
for type-checking usage type constructs
(see~Figure~\ref{fig:typing-classes}). After type-checking $u$, the
object fields in $\Sigma$ are a subset of the original ones, that have
been consumed, inside the object, to the end. Those which are missing
have been passed as parameters, or returned from methods, and thus
removed from the environment. This means that the object fields in a
class final environment have always unrestricted types, either an
unrestricted recursive type or $\un~\End$. Function $\un(\Sigma)$
recursively calls predicate $q(t)$, passing each field type in
$\Sigma$ as argument, in order to guarantee that no linear fields
exist in the class final environment. We also define $q(\Sigma)$ if 
and only if $o.f: t \in \Sigma$ implies $q(t)$, so that the type system 
can check that no linear fields are allowed in shared objects.

\begin{figure}[t]
\centering\setlength\lineskip{\typingRuleSkip}
\hfill\\
\AxiomC{$\forall i \in I \qquad 
   s_i~\method{t_i}{m_i}{t_i'~x_i}{e_i} \in C.\methodsterm$}  
\AxiomC{$q(\Sigma)$ }
\AxiomC{\judgment \Sigma,x_i : t_i'  > e_i : t_i < \Gamma /}
\noLine
\TrinaryInfC{$x_i : t_i'' \in \Gamma \Rightarrow \un(t_i'')$ 
   \quad $\Gamma = \choicem{\_}{\_} \Rightarrow t_i = \booleantype$
   \quad \judgmentu \Theta; \Gamma{\setminus x_i} > C u_i < \Gamma'/}
\rulename{T-Branch}
\UnaryInfC{\judgmentu \Theta; \Sigma > C q~\branchm{m_i}{u_i}{} < \Gamma'/}
\DisplayProof
\hfil
\AxiomC{\judgmentu \Theta; \Gamma' > C u_t < \Gamma/}
\AxiomC{\judgmentu \Theta; \Gamma'' > C u_f < \Gamma/}
\rulename{T-Variant}
\BinaryInfC{\judgmentu \Theta; \choicem{\Gamma'}{\Gamma''} > C \choicem{u_t}{u_f} < \Gamma/}
\DisplayProof\\
\hfil
\axiomname{T-UsageVar}
\judgmentu (\Theta, X : \Gamma); \Gamma > C X < \Gamma' /
\hfil
\AxiomC{\judgmentu (\Theta, X : \Gamma); \Gamma > C u < \Gamma'/}
\rulename{T-Rec}
\UnaryInfC{\judgmentu \Theta; \Gamma > C \mu X.u < \Gamma'/}
\DisplayProof
\hfil
\caption{Typing rules for classes}
\label{fig:typing-classes}
\end{figure}

%%% Local Variables: 
%%% mode: latex
%%% TeX-master: "main"
%%% End: 

The rules in Figure~\ref{fig:typing-classes} describe how the four 
usage constructs (\textit{cf.}~Figure~\ref{fig:syntax}) are typed in $\mool$. 
{\textsc{T-Branch} conducts the type-checking of the methods defined in the class 
usage type in the order in which they appear in the specified sequence.
The appropriate rules for expressions (see 
Figures~\ref{fig:typing-exps}~and~\ref{fig:typing-exps-calls}) are called 
by \textsc{T-Branch} so that each method body $e_i$ can be type-checked. The initial 
environment is a single map of fields and parameters, denoted by $\Sigma$, while the 
final one may be a pair of maps if the method $m_i$ returns a value of type 
$\booleantype$. Thus $\Gamma$ is used instead. The relation $q \subset q'$ is reflexive and 
transitive, and in particular $\lin \subset \un$. When exiting the method body, 
the parameter type in $\Gamma$ may have changed from the initial type $t_i'$ to the 
final one $t_i''$. Additionally, the rule 
requires that $t_i''$ is unrestricted, so the function $\un(t_i'')$ checks that the 
parameter has been consumed to the end. Finally, the parameter entry must be removed 
from $\Gamma$ before the rule advances to type-checking the method continuations~$u_i$, 
where again the field typing may change, modifying the initial 
environment $\Gamma$ to the final one $\Gamma'$. 
 
In checking variant types, \textsc{T-Variant} requires a consistency
between the variants final environment $\Gamma$, after passing
$\Gamma'$, the left environment, to type the $\trueterm$ variant, and
$\Gamma''$, the right environment, to type the $\falseterm$
variant. \textsc{T-UsageVar} simply reads the type variable $X$ from
map~$\Theta$. Because $X$ represents an infinite branch in the usage
type tree that no program can ever consume to the end, the rule cannot
define the final environment as being the same as the initial one,
which means that the final environment can be whichever we
want. \textsc{T-Rec}, not only reads the type from $\Theta$, but
also checks the type against the class, using \textsc{T-UsageVar} to
type usage variables $X$, and the other two rules, \textsc{T-Branch}
and \textsc{T-Variant}, to type branch and variant constructs. The
final environment $\Gamma'$ may be different, because field types may
have changed, after type-checking expressions in method bodies (recall
that \textsc{T-Branch} calls the appropriate rules for expressions).

We now present the rules for expressions. In order to make them more readable, we 
use the single environment $\Sigma$ everywhere a map is needed to index a field or a 
parameter; otherwise the more general $\Gamma$ is used. Figures~\ref{fig:typing-exps} 
and \ref{fig:typing-exps-calls} define the rules for the remaining expressions of the 
top level language. 

\begin{figure}[t]
\centering\setlength\lineskip{\typingRuleSkip}
\AxiomC{$\lin(t)$}
\axiomname{T-LinVar}
\UnaryInfC{\judgment \Sigma, x : t > x : t < \Sigma /}
\DisplayProof\hfil
\AxiomC{$\un(t)$}
\rulename{T-UnVar}
\UnaryInfC{\judgment \Sigma, x : t > x : t < \Sigma, x : t/}
\DisplayProof\hfil
\AxiomC{$\lin(t)$}
\axiomname{T-LinField}
\UnaryInfC{\judgment \Sigma, o.f : t > o.f : t < \Sigma /}
\DisplayProof\hfil
\AxiomC{$\un(t)$}
\rulename{T-UnField}
\UnaryInfC{\judgment \Sigma, o.f : t > o.f : t < \Sigma, o.f : t/}
\DisplayProof\hfil
\AxiomC{\judgment \Gamma > e : t < \Gamma'' /}
\AxiomC{\judgment \Gamma'' > e' : t' < \Gamma' /}
\rulename{T-Seq}
\BinaryInfC{\judgment \Gamma > \seq{e}{e'} : t' < \Gamma' /}
\DisplayProof
\hfill
\AxiomC{\judgment \Sigma > e : t < \Sigma' /}
\AxiomC{$\Sigma'(o.f) = t$}
\AxiomC{$\un(t)$}
\rulename{T-Assign}
\TrinaryInfC{\judgment \Sigma > \assign{o.f}{e} : \unit < \Sigma' /}
\DisplayProof\hfil
\axiomname{T-New}
\judgment \Gamma > \new C : \objecttype{C}{C.\usageterm} < \Gamma /
\hfil
\AxiomC{\judgment \Gamma > e : t < \Gamma' /}
\AxiomC{$\un(t)$}
\rulename{T-Spawn}
\BinaryInfC{\judgment\Gamma > \spawn~e : \unit < \Gamma' /}
\DisplayProof
\hfil
\caption{Typing rules for simple expressions}
\label{fig:typing-exps}
\end{figure}

%%% Local Variables: 
%%% mode: latex
%%% TeX-master: "main"
%%% End: 

\begin{figure}[t]
\centering\setlength\lineskip{\typingRuleSkip}
\AxiomC{\judgment \Gamma > e : t < \Sigma /}
\AxiomC{$\Sigma(o.f) = \objecttype C{\_~\branchm{m_i}{u_i}{i\in I}}$}
\noLine
\BinaryInfC{$j\in I\qquad \method{t}{m_j}{t\,x}{\_}\in C.\methodsterm$}
\rulename{T-Call}
\UnaryInfC{\judgment \Gamma > \methcal{o.f}{m_j}{e} : t <
  \changetype{\Sigma}{o.f}{\objecttype{C}{u_j}} /}
\DisplayProof\hfil
\AxiomC{\judgment \Gamma > \methcal{o.f}{m}{e} : \booleantype < \Sigma /}
\noLine
\UnaryInfC{\judgment\changetype{\Sigma}{o.f}{\objecttype{C}{u_t}} > 
   e' : t < \Gamma' /}
\AxiomC{$\Sigma(o.f) = \objecttype C{\choicem{u_t}{u_f}}$}
\noLine
\UnaryInfC{\judgment\changetype{\Sigma}{o.f}{\objecttype{C}{u_f}} > e'' : t < 
  \Gamma' /}
\rulename{T-IfV}
\BinaryInfC{\judgment \Gamma > \ifelse{\methcal{o.f}{m}{e}}{e'}{e''} : t < 
  \Gamma' /}  
\DisplayProof\hfil
\AxiomC{\judgment \Gamma > \methcal{o.f}{m}{e} : \booleantype < \Sigma /}
\noLine
\UnaryInfC{$\Sigma(o.f) = \objecttype C{\choicem{u_t}{u_f}}$ \qquad
   \judgment\changetype{\Sigma}{o.f}{\objecttype{C}{u_t}} > e' : t < \Gamma /}
\rulename{T-WhileV}
\UnaryInfC{\judgment \Gamma > \while{\methcal{o.f}{m}{e}}{e'} : \unit < 
   \changetype{\Sigma}{o.f}{\objecttype{C}{u_f}} /}  
\DisplayProof\hfil
\AxiomC{\judgment\Gamma > e : \booleantype < \Gamma' /}
\AxiomC{\judgment\Gamma' > e' : t < \Gamma'' /}
\AxiomC{\judgment\Gamma' > e'' : t < \Gamma'' /}
\rulename{T-If}
\TrinaryInfC{\judgment \Gamma > \ifelse{e}{e'}{e''} : t < \Gamma'' /}
\DisplayProof\hfil
\AxiomC{\judgment\Gamma > e : \booleantype < \Gamma' /}
\AxiomC{\judgment\Gamma' > e' : t < \Gamma /}
\rulename{T-While}
\BinaryInfC{\judgment \Gamma > \while{e}{e'}: \unit < \Gamma' /}
\DisplayProof\hfil\\
\AxiomC{\judgment \Gamma > e : t < \Gamma' /}
\rulename{T-InjL}
\UnaryInfC{\judgment \Gamma > e : t < \choicem{\Gamma'}{\Gamma''} /}  
\DisplayProof\hfil
\AxiomC{\judgment \Gamma > e : t < \Gamma'' /}
\rulename{T-InjR}
\UnaryInfC{\judgment \Gamma > e : t < \choicem{\Gamma'}{\Gamma''} /} 
\DisplayProof\hfil\\
\AxiomC{\judgment\Gamma > e : C[u] < \Gamma' /}
\AxiomC{$C[u]\subt C[u']$}
\rulename{T-Sub}
\BinaryInfC{\judgment\Gamma > e : C[u'] < \Gamma' /}
\DisplayProof\hfil
\AxiomC{\judgment\Gamma > e : t < \Sigma /}
\AxiomC{$\Sigma\subt\Sigma'$}
\rulename{T-SubEnv}
\BinaryInfC{\judgment\Gamma > e : t < \Sigma' /}
\DisplayProof\hfil
\caption{Typing rules for calls and control flow expressions}
\label{fig:typing-exps-calls}
\end{figure}

%%% Local Variables: 
%%% mode: latex
%%% TeX-master: "main"
%%% End: 

\textsc{T-LinVar} and \textsc{T-UnVar} evaluate the types of method parameters, 
distinguishing linear from unrestricted ones. Recall that, as we cannot call 
methods on parameters, they must be first assigned to fields. In the case of 
parameters with linear types, the type system requires that they are removed 
from the environment, so that they can no longer be used after assignment. 
Predicates $\lin(t)$ and $\un(t)$ are used to determine the status of a given 
type. In practice, primitive types ($\unit$ and $\booleantype$) are always 
unrestricted; for object types the current status varies with the type. 
\textsc{T-LinField, T-UnField} are similar to the rules defined above 
for parameters, removing fields with linear types from the environment.

Typing the sequential composition of expressions is simple: 
\textsc{T-Seq} considers the typing of the second subexpression taking into 
account the effects of the first one on the environment, which may be different 
when $e$ represents a method call or a control flow expression. \textsc{T-Assign} 
formalises assignments to fields. The type of the reference and the expression must 
be consistent and, because of linearity, it must be unrestricted. The type of the 
entire assignment expression is $\unit$ (as opposed to $t$) to enforce that a linear 
reference goes out of scope after appearing on the right-hand side of the assignment. 
\textsc{T-New} simply states that a new object has the initial usage type declared by 
its class. \textsc{T-Spawn} requires not only that the thread body is typable, but 
also that it is not of a linear type; otherwise one could create a linear reference 
and not use it to the end (suppose we wrote \lstinline|spawn new C()|). The type of 
the entire $\spawnterm$ expression is $\unit$, because the evaluation is performed 
only for its effect (creating a new thread); no result is ever returned. 

\textsc{T-Call} requires that the receiver has an appropriate usage type. The call 
results in the receiver changing its type to $C[u_j]$, with the final environment 
$\Sigma$ being updated accordingly, while at the same time the type advances 
from~$\branchm{m_i}{u_i}{}$ to $u_j$.

As opposed to the rules presented thus far, the remaining ones in 
Figure~\ref{fig:typing-exps-calls} are not syntax-directed, which means that to 
implement them additional information is needed as they are evaluated. To simplify 
the rules, for an object to have a variant type, the method call on which the variant 
type depends must be performed on the condition of a control flow expression (and not 
on an arbitrary assignment). Recall that a variant type is tied to the result of a 
method and, depending on the value returned, an object type may be modified 
differently. The type system requires that the value is tested on the condition of 
the expression, so as to have the type immediately deduced. We believe that this 
restriction does not impair current object-oriented programming practices, 
as it reflects how a variant type is typically given to an object.

\textsc{T-IfV} and \textsc{T-WhileV} are particular cases of the more general rules 
for control flow expressions that we describe below. The order in which we present 
them here corresponds to the actual order in which the type-checker evaluates them. 
Both rules depend on $\textsc{T-Call}$ to deduce the return type of the method tested 
on the condition. In the case of $\textsc{T-IfV}$, both branches use $\Sigma$, the 
environment that results from the call, as their initial environment. The method 
returned value defines how $o.f$ type is updated, thus determining which branch (the 
$\thenterm$ or the $\elseterm$) shall be executed. However, because only one of the 
branches can be executed, the rule enforces that the two branches should be 
consistent, sharing the same final environment $\Gamma'$. \textsc{T-WhileV} follows a 
similar pattern: a $\whileterm$ expression can be reduced to an $\ifterm$ that is 
repeated while a particular condition returns $\trueterm$. The type of the entire 
loop expression is $\unit$, because its result can never be used. 
\textsc{T-If, T-While} are standard rules for control flow expressions, and should be 
used only after the above rules have been tried.

\textsc{T-InjL} and \textsc{T-InjR} build a variant environment as follows: an 
environment $\Gamma$ becomes $\choicem{\Gamma}{\Gamma'}$ by injecting $\Gamma$ 
on the left using rule~\textsc{T-InjL}, and becomes $\choicem{\Gamma'}{\Gamma}$ 
by injecting on the right using rule~\textsc{T-InjR}. Typically, these rules 
build the final environment of a $\booleantype$ method to which a variant is tied. 

\textsc{T-Sub} and \textsc{T-SubEnv} are similar to the subsumption rules 
described for the language in modular session 
types~\cite{gay.vasconcelos.etal_modular-session-types}. \textsc{T-Sub} is a standard 
subsumption rule that simply says that whenever we can prove that type $t'$ is a 
subtype of $t$, we can treat $t'$ as if it were type $t$. \textsc{T-SubEnv} allows 
subsumption in the final environment, and is used for the branches of control flow 
expressions.

%%% Local Variables: 
%%% mode: latex
%%% TeX-master: "main"
%%% End: 

%%% Local Variables: 
%%% mode: latex
%%% TeX-master: "main"
%%% End: 

\section{Concluding Remarks}
\label{sec:conclusion}

Session types, introduced
in~\cite{HondaK:typdi,HondaK:lanptd,HondaK:intblt}, have been proposed
to enhance the verification of programs at compile-time by specifying
the sequences and types of messages in communication protocols.
Traditionally associated with communication channels, session types
provide a means to enforce that channel implementations obey the
requirements stated by their types. Originally developed for dyadic
sessions, the concept was then extended to multi-party
sessions~\cite{HondaK:mulast}. Programming languages that implement
session types come in all flavours: pi calculus, an idealised
concurrent programming language in the context of which the original
concepts were developed, functional languages~\cite{NeubauerM:impst,
  VasconcelosV:session-types-functional, VasconcelosVT:typmfl},
CORBA~\cite{VallecilloVasconcelosEtal:typing-behavior-objects},
object-oriented languages~\cite{DezaniCiancagliniM:sestoo,
  DezaniCiancagliniM:disool,
  gay.vasconcelos.etal_modular-session-types,
  Mostrous05moose}. Channels as conceived in session type theory are
special entities that carry messages of different types,
bi-directionally, in a specific sequence between two end points.
These channels are usually implemented in a socket-like style, and, to
our knowledge, none of the previous attempts to integrate session
types into object-oriented programming ever abstracted this notion of
communication channels. The work on
Moose~\cite{DezaniCiancagliniM:sestoo,DezaniCiancagliniM:disool}, a
multi-threaded object-oriented calculus with session types, was the
first attempt to marriage the (concurrent) object-oriented paradigm
and session types. Unlike our work, this and subsequent work have kept
distinct mechanisms for local and remote communication, in the form of
method call and channel operations, respectively.

In the literature, several lines of research can be found that reveal 
similarities with session type theory. One of these lines introduces the concept
of typestate \cite{StromYemini:typestates} in which the state of the object in 
some particular context determines the set of available operations in that 
context, based on pre- and post-conditions. Objects, by nature, can be in 
different states throughout their life cycle. The concept involves static 
analysis of programs at compile-time so that all the possible states of an 
object and associated legal operations can be tracked at each point in the 
program text. Typestate checking has been incorporated in several programming 
languages \cite{BierhoffAldrich:plural, DeLine03thefugue, FahndrichM:lansfr}, 
and some ideas relate very closely to session type recent approach on 
modularity.   

This paper formalises a mini class-based language that uses primitives
consistent with most object-oriented languages, and incorporates
support for the specification and static checking of usage
protocols. The existing work on modular session
types~\cite{gay.vasconcelos.etal_modular-session-types} has been the
inspiration for our specification language and type system, where we
replace operations on communication channels by remote method
invocations. Another distinguishing feature comes from allowing
objects to change from a linear status to a shared one. We also use a
standard synchronization mechanism to control concurrency, instead of
the well-known shared channel on which linear channels are created. 
We have designed a simple operational semantics and a static 
type system which checks client code conformance to method call 
sequences and behaviourally constrained aliasing, as expressed 
by programmers in class usage types.
% Environments are used to track consumption of linear references.

One of the main limitations that can be pointed out to the technical discussion 
in this paper is that it does not cover a full formal treatment of the
$\mool$ language, nor does it present proofs. However, we are confident that, 
with minor adjustments, we can adopt the techniques used in the work on modular
session types~\cite{gay.vasconcelos.etal_modular-session-types} and
apply them to our language.

A feature which our system does not consider is the treatment of exceptions.
We do not predict any sort of mechanism for letting a method throw a checked 
exception, so that a client can execute some predefined error-handling 
code when it happens. However, Java-like error-management techniques, 
or alternatively, some exceptional default state encoded in the usage 
type~(see \cite{MilitaoF:yak}) could be easily added to our language.

Finally, although we tried to introduce some flexibility in our approach 
to linearity by letting linear objects evolve into shared ones, 
we still have not quite found a middle ground: our language 
entirely bans the aliasing of linear types, while completely allowing
the aliasing of unrestricted types. Allowing limited forms of aliasing
without loosing track of an object state is a topic that has occupied many 
researchers, and which we also plan to pursue in future work.

%%% Local Variables: 
%%% mode: latex
%%% TeX-master: "main"
%%% End: 

\paragraph{Acknowledgements}

This work was funded by project ``Assertion-Types for Object-Oriented
Programming'', FCT (PTDC/EIA-CCO/105359/2008).

%%% Local Variables: 
%%% mode: latex
%%% TeX-master: "main"
%%% End: 

\nocite{*}
\bibliographystyle{eptcs}
\bibliography{main}

\end{document}